\documentclass[traditabstract]{aa}
\bibpunct{(}{)}{;}{a}{}{,} % to follow the A&A style
\usepackage{graphicx}
\usepackage[varg]{txfonts}
\usepackage{color}
\usepackage[utf8]{inputenc}
\usepackage{hyperref}
\begin{document}

\title{Pre-perihelion evolution of the NiI/FeI abundance ratio in the coma of the interstellar comet 3I/ATLAS. From extreme to normal\thanks{Based on observations made with the ESO Very Large Telescope at the Paranal Observatory under programs 115.C-0282, 115.C-2317, and 2115.C-5030.}}
\author{Damien Hutsem\'ekers\inst{1},
        Jean Manfroid\inst{1},
        Emmanu\"el Jehin\inst{1},
        Cyrielle Opitom\inst{2},
        Rohan Rahatgaonkar\inst{4},
        Michele Bannister\inst{3},
        Juan Pablo Carvajal\inst{4},
        Rosemary Dorsey\inst{5},
        K.~Aravind\inst{1},
        Baltasar Luco\inst{4},
        Brian Murphy\inst{2},
        Thomas H. Puzia\inst{4}
        }
\institute{
    STAR, Institut d'Astrophysique et de G\'eophysique, Universit\'e de Li\`ege, All\'ee du 6 Ao\^ut 19c, 4000 Li\`ege, Belgium
    \and
    Institute for Astronomy, University of Edinburgh, Royal Observatory, Edinburgh EH9 3HJ, United Kingdom
    \and
    School of Physical and Chemical Sciences - Te Kura Mat\={u}, University of Canterbury, Private Bag 4800, Christchurch 8140, New Zealand
    \and
    Institute of Astrophysics, Pontificia Universidad Cat\'olica de Chile, Av.~Vicu\~na Mackenna 4860, 7820436 Macul, Santiago, Chile
    \and
    Department of Physics, University of Helsinki, P.O. Box 64, 00014 Helsinki, Finland
    }

\date{Received ; accepted: }
\titlerunning{NiI/FeI in 3I/ATLAS} 
\authorrunning{D. Hutsem\'ekers et al.}
\abstract{Emission lines of FeI and NiI are commonly found in the coma of Solar System comets, even at large heliocentric distances. These atoms are  most likely released from the surface of the comet's nucleus or from a short-lived parent. The presence of these lines in cometary spectra is unexpected because the surface blackbody equilibrium temperature is too low to allow the sublimation of refractory minerals containing these metals. These lines were also found in the interstellar comet 2I/Borisov, which has a NiI/FeI abundance ratio similar to that observed in Solar System comets. On average, this ratio is one order of magnitude higher than the solar Ni/Fe abundance ratio. Here, we report observations of the interstellar comet 3I/ATLAS, which were carried out with the ESO Very Large Telescope equipped with the UVES and X-shooter spectrographs. Spectra were obtained at heliocentric distances ranging from 3.14 to 1.85 au. NiI was detected at all epochs. FeI was only detected at heliocentric distances smaller than 2.64~au.  We estimated the NiI and FeI production rates by comparing the observed line intensities with those produced by a dedicated fluorescence model. Comet 3I first exhibited extreme and unusual NiI/FeI abundance ratios during the initial stages of its activity. However, as its heliocentric distance decreased, this ratio became indistinguishable from those observed in Solar System comets and in comet 2I/Borisov. Comet 3I was found to be C$_2$-depleted, with a NiI/FeI abundance ratio finally consistent with other C$_2$-depleted comets. Nevertheless, comet 3I remains exceptional due to its high, total production rate of NiI and FeI, which is at least one order of magnitude larger than that of other comets. We interpreted these observations assuming that the NiI and FeI atoms were released through the sublimation of Ni(CO)$_4$ and Fe(CO)$_5$ carbonyls. This scenario provides a straightforward explanation for the asymmetric release of NiI and FeI atoms in the cometary coma and how it depends on the heliocentric distance. It also supports the presence of carbonyls in the cometary material.}
\maketitle
%
%
%________________________________________________________________
%

\section{Introduction}
\label{sec:intro}

Numerous FeI and NiI emission lines have been identified in the spectra of about 20 Solar System comets observed in the last two decades at heliocentric distances ($r_h$) ranging from 0.68 to 3.25~au \citep{Manfroid2021,Hutsemekers2021,Hmiddouch2025}.  The presence of these lines in cometary spectra at such distances from the Sun was unexpected because the surface blackbody equilibrium temperature is too low to allow the sublimation of silicate and sulfide minerals containing the metals.   Furthermore, the mean abundance ratio NiI/FeI was found to be one order of magnitude higher than the solar ratio, and higher than the ratios estimated in the dust of 1P/Halley \citep{Jessberger1988} and in the coma of the Sun-grazing comet Ikeya-Seki \citep{Manfroid2021}. \citet{Manfroid2021} advanced several mechanisms to explain these observations, in particular superheating of Ni-rich sulfides, possibly located in nanoparticles, and sublimation of organometallic complexes, such as carbonyls (see also \citealt{Bromley2021}  and \citealt{Rahatgaonkar2025}).

Interstellar comets, with potentially different chemical compositions, offer a unique opportunity to better understand the origin of the NiI and FeI atoms observed in cometary comae. FeI and NiI were found in the interstellar comet 2I/Borisov (hereafter 2I) with a NiI/FeI abundance ratio similar to the Solar System comets \citep{Guzik2021,Opitom2021}. NiI was recently found in the interstellar comet 3I/ATLAS (hereafter 3I) at a heliocentric distance of 3.88~au, before the onset of CN and FeI \citep{Rahatgaonkar2025}. This could support an unusual composition of comet 3I, as is suggested by the very high CO$_2$/H$_2$O abundance ratio found by \citet{Cordiner2025}.

In this work, we report the first detection of FeI emission lines in the coma of comet 3I at a heliocentric distance of 2.64~au. We measured the NiI/FeI abundance ratio at different heliocentric distances and compared it to the ratio observed in Solar System comets and comet 2I to determine whether comet 3I is unique. We show that the release of NiI and FeI atoms at large distances from the Sun can be explained by the sublimation of short-lived species, possibly carbonyls.

\section{Observations and data reduction}
\label{sec:obs}

Observations were carried out from August 12 to September 14, 2025, with the Very Large Telescope (VLT) at the European Southern Observatory (ESO), equipped with the UV-Visual Echelle Spectrograph (UVES\footnote{UVES User Manual, VLT-MAN-ESO-13200-1825, \\ https://www.eso.org/sci/facilities/paranal/instruments.html}). The standard settings 346+580 (dichroic 1), 390+580 (dichroic 1), and 437+860 (dichroic 2) were used. The  blue settings centered at 346~nm, 390~nm, and 437~nm cover the spectral ranges 3030-3880~\AA, 3260-4540~\AA, and 3730-4990~\AA, respectively. A nonstandard setting centered at 348~nm, corresponding to the spectral range 3100-3900~\AA, was also used to simultaneously cover the regions containing the OH and CN bands. Most of the time, a 1.8\arcsec-wide slit was used, delivering a resolving power of about 35000. The observing circumstances are summarized in Table~\ref{tab:obs1}. Raw frames were first processed to remove cosmic ray hits using the Python implementation of the ``lacosmic'' package \citep{VanDokkum2001,VanDokkum2012}. The data were reduced with the UVES pipeline\footnote{UVES Pipeline User Manual, VLT-MAN-ESO-19500-2965,\\ https://www.eso.org/sci/software.html} to obtain wavelength- and flux-calibrated two-dimensional spectra. One-dimensional spectra were then extracted by integrating over the full slit length (Table~\ref{tab:data1}), using custom procedures. The scattered spectrum of the Sun (dust, twilight) was removed in the manner described in \citet{Manfroid2009}.

We also considered the spectra obtained with the VLT X-shooter spectrograph\footnote{X-shooter User Manual, ESO-270545, \\ https://www.eso.org/sci/facilities/paranal/instruments.html} \citep{Vernet2011} and described in detail by \citet{Rahatgaonkar2025}, in particular the UVB spectra obtained from July 23 to August 21, 2025, which contain the first detections of NiI lines in comet 3I. These observations were secured with a 1.6\arcsec-wide slit, providing a resolving power of about 3200. Another series of spectra, containing both the NiI and FeI lines, was obtained from September 20 to 25, 2025, with the same instrumental settings. The observing circumstances are summarized in Table~\ref{tab:obs2}. All of these spectra were independently reduced in the present work.  The X-shooter pipeline\footnote{X-shooter Pipeline User Manual, VLT-MAN-ESO-14650-4840,\\ https://www.eso.org/sci/software.html} was used to obtain wavelength- and flux- calibrated two-dimensional spectra. Since the target was not always positioned in the middle of the 10\arcsec -long slit, the one-dimensional spectra were extracted over a shorter, 7.5\arcsec -long slit centered on the target. The scattered spectrum of the Sun was subtracted using the solar analog star HD150469, which was observed with the same settings as the comet \citep{Rahatgaonkar2025}.

\section{FeI line detection and production rate measurements}
\label{sec:measurements}

\begin{figure}[]
\centering
\resizebox{0.95\hsize}{!}{\includegraphics*{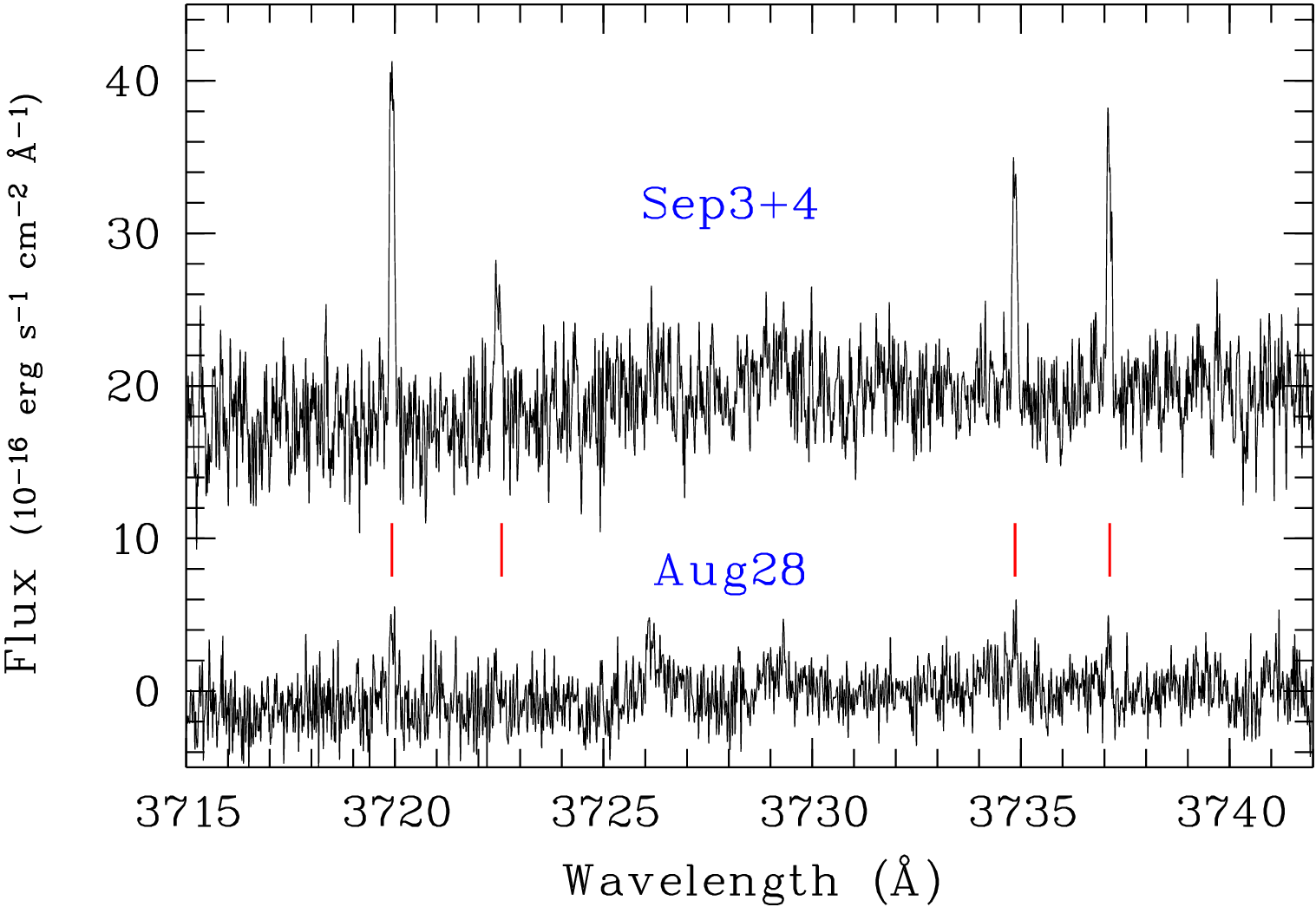}}
\caption{Continuum-subtracted spectra of comet 3I obtained on August 28 and September 3+4 with UVES. Three narrow FeI lines were detected on August 28 in this spectral range, FeI $\lambda$3719.93\AA\ being the brightest iron line detected in our spectra. In the average spectrum obtained in September 3+4 (shifted vertically for clarity), the FeI lines are brighter, and four of them were detected (as is indicated by the vertical red lines). Broad emission features due to background [OII] emission at 3727\AA\ and 3729\AA\ are also observed.}
\label{fig:spec}
\end{figure}

\begin{figure}[]
\centering
\resizebox{0.95\hsize}{!}{\includegraphics*{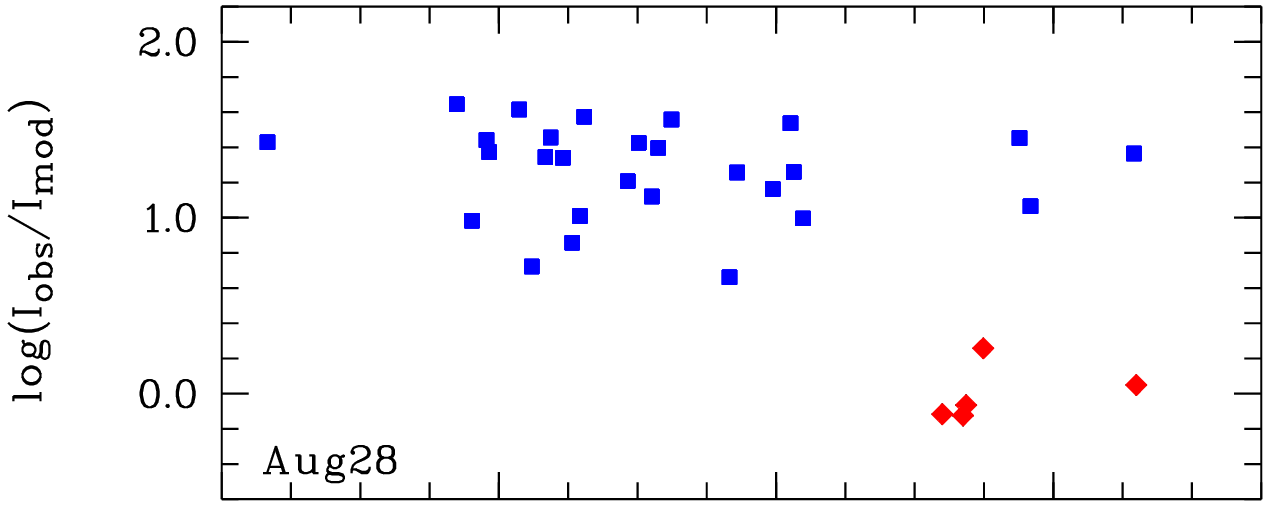}}\\
\resizebox{0.95\hsize}{!}{\includegraphics*{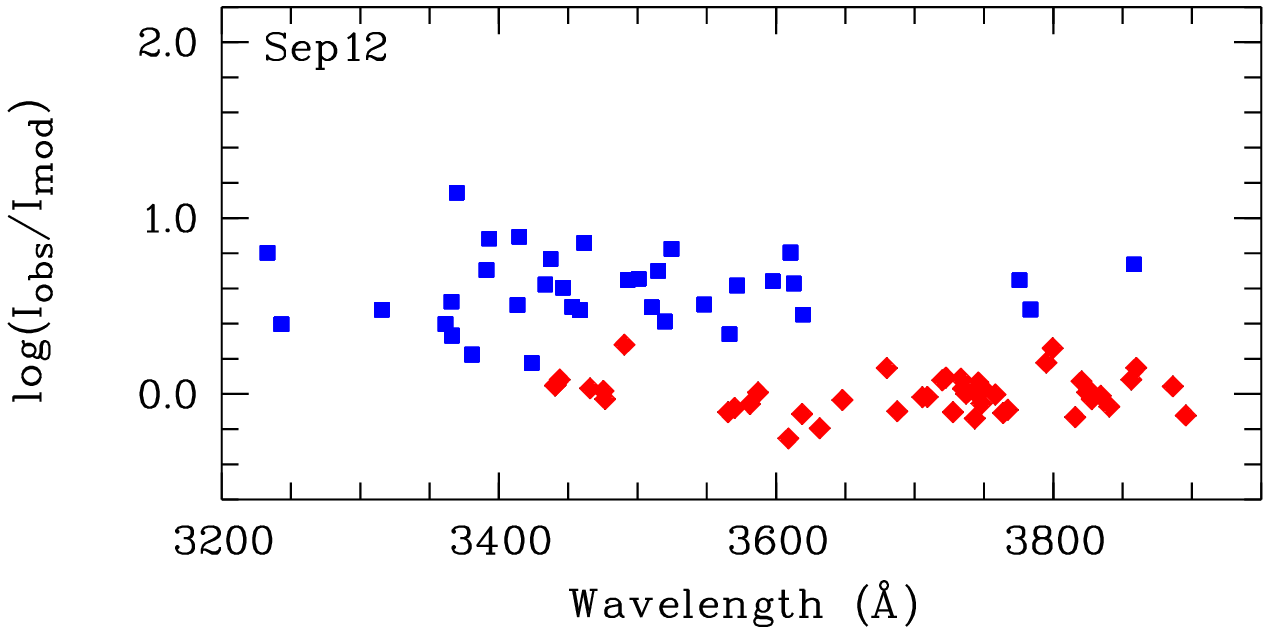}}
\caption{Ratio $\log_{10} (I_{\rm obs} / I_{\rm mod})$ for the FeI (red diamonds) and NiI (blue squares) lines measured in comet 3I for observations secured with UVES on August 28 (top) and September 12 (bottom). $I_{\rm obs}$ represents the observed line intensities and $I_{\rm mod}$ the intensities computed with the fluorescence model. $I_{\rm obs} / I_{\rm mod}$ is proportional to the column density of the atoms \citep{Manfroid2021}. The separation between the mean values computed for NiI and FeI gives the NiI/FeI abundance ratio. The ratios have been shifted on the y axis so that the mean value of $\log_{10} (I_{\rm obs} / I_{\rm mod})$ is zero for FeI.  }
\label{fig:model}
\end{figure}

\begin{figure}[]
\centering
\resizebox{0.95\hsize}{!}{\includegraphics*{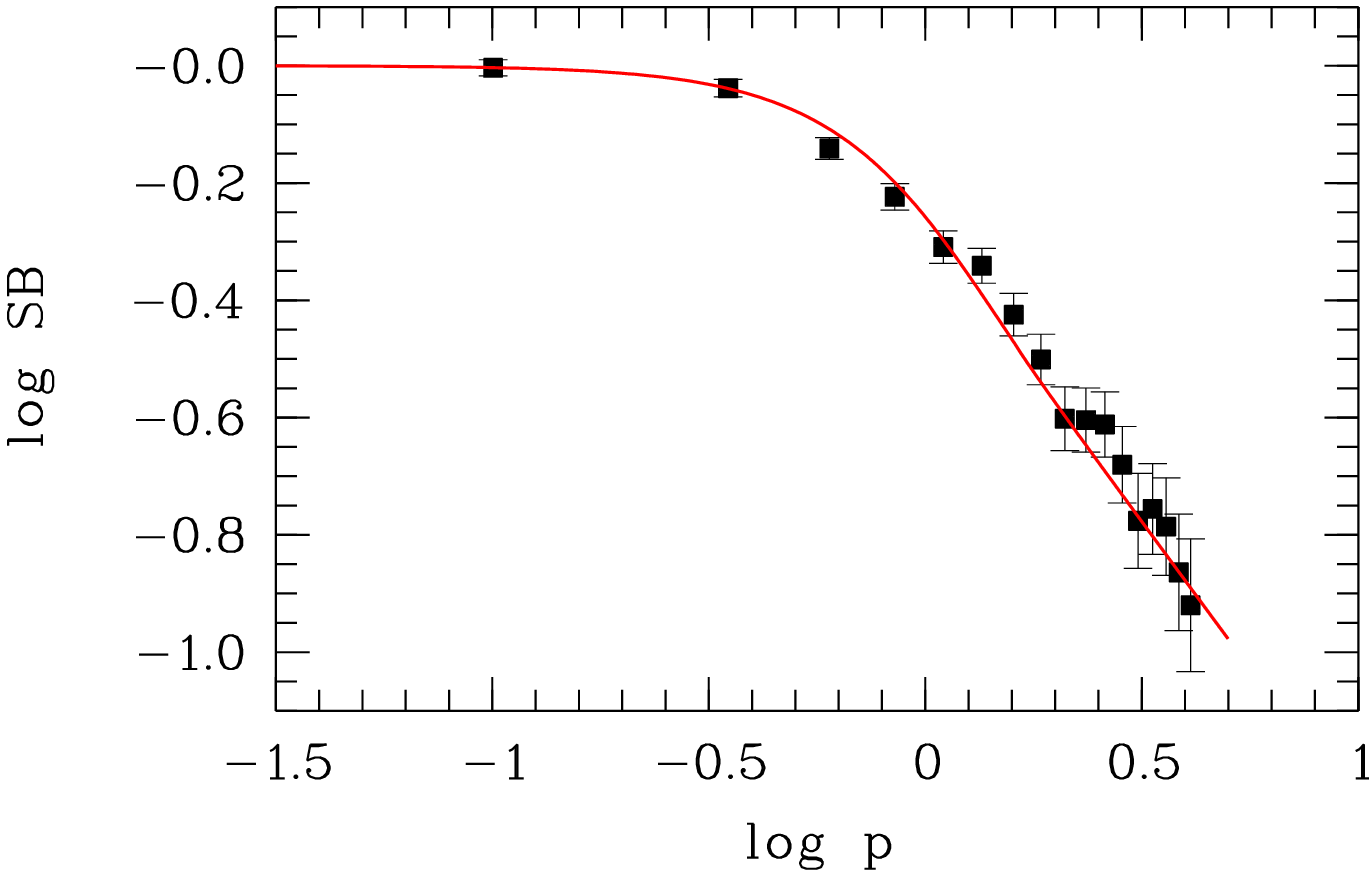}}
\caption{Spatial profile of the brightest NiI line ($\lambda$ 3458\AA) observed with UVES on September 4. The measured surface brightness (SB; normalized to one at the photocenter) is plotted as a function of the projected nucleocentric distance, $p$, in arcsec. The red line represents SB $\propto p^{-1}$ convolved with a 1.5\arcsec\ full-width-at-half-maximum Gaussian to account for the seeing and tracking imperfections.}
\label{fig:rp}
\end{figure}

While NiI lines were observed in the spectra obtained before August~28 and detected as far as 3.88 au \citep[][see also Table~\ref{tab:data1}]{Rahatgaonkar2025}, FeI lines were only detected in the spectra obtained on August~28 with UVES, and after that date when the comet was at $r_h \leq 2.64$ au. Some of the brightest FeI lines are shown in Fig.~\ref{fig:spec}. After August 28, the lines steadily intensified, and new ones emerged (Figs.~\ref{fig:spec_ni} and~\ref{fig:spec_fe}).

The intensities of unblended FeI and NiI emission lines were measured in the UVES and X-shooter spectra for each epoch, and compared to the intensities computed with a dedicated fluorescence model to derive the FeI and NiI column densities. This model, described in detail in \citet{Manfroid2021}, considers a large number of transitions and atomic levels for both the FeI and NiI atoms. Since FeI and NiI lines are also present in absorption in the solar light that irradiates the cometary atmosphere, the excitation can strongly depend on the Doppler shift between the comet and the solar spectra. Therefore, the high-resolution structure of the solar spectrum and the relative velocity between the comet and the Sun must be considered.

A comparison of observed and computed line intensities for FeI and NiI is shown in Fig.~\ref{fig:model}, for two epochs. The ratio $I_{\rm obs}/I_{\rm mod}$ gives the atomic column density  \citep{Manfroid2021}. Although there is clearly some dispersion between the individual measurements, the average of the intensity ratios is well defined for both atomic species. The difference of the average values gives the NiI/FeI abundance ratio, which is clearly higher on August 28 than on September 12 (see also Table~\ref{tab:data1}). 

One important characteristic of the FeI and NiI emission lines is their short spatial extent. In spectra obtained for comets 103P/Hartley2 and 46P at geocentric distances of 0.17 and 0.09 au, respectively, the surface brightness was found to be inversely proportional to the projected distance from the nucleus, $p$ \citep{Manfroid2021,Hutsemekers2021}. Such a spatial profile indicates that the FeI and NiI atoms originate at small nucleocentric distances, either directly ejected from the surface of the nucleus or released from a short-lived parent. As is shown in Fig.~\ref{fig:rp}, a similar profile is observed for comet 3I. 

Assuming that the emission line surface brightness decreases as $p^{-1}$ for both NiI and FeI, production rates were computed from the column densities according to the formulae given in \citet{Manfroid2021}. We assumed a constant ejection velocity, $v$ = 0.85 $r_h^{-0.5}$ km s$^{-1}$ \citep{Cochran1993}. The resulting production rates and abundance ratios are given in Table~\ref{tab:data1}, together with the number of lines used in the analysis. The NiI production rates measured for the first X-shooter dataset are in good agreement with those reported in \citet{Rahatgaonkar2025}, although they are slightly smaller due a more conservative measurement of the line intensities.  Since NiI and FeI were simultaneously observed, abundance ratios can be derived from either column densities or production rates.

The production rates of OH, CN, and C$_2$ were also estimated as in \citet{Manfroid2021}, using the OH(0-0) band at 3090~\AA, the CN(0-0) band at 3870~\AA, and the C$_2$ band at 5140~\AA. The rates were derived using the fluorescence efficiencies and Haser scalelengths from \citet{Schleicher1988}, \citet{Cochran1993}, \citet{AHearn1995}, and \citet{Schleicher2010}, with a parent and daughter velocity of 0.85 $r_h^{-0.5}$ km s$^{-1}$ \citep{Cochran1993}. These production rates are given in Table~\ref{tab:data2}.

\section{Comparison of comet 3I to other comets}
\label{sec:results}

\begin{figure}[]
\centering
\resizebox{0.95\hsize}{!}{\includegraphics*{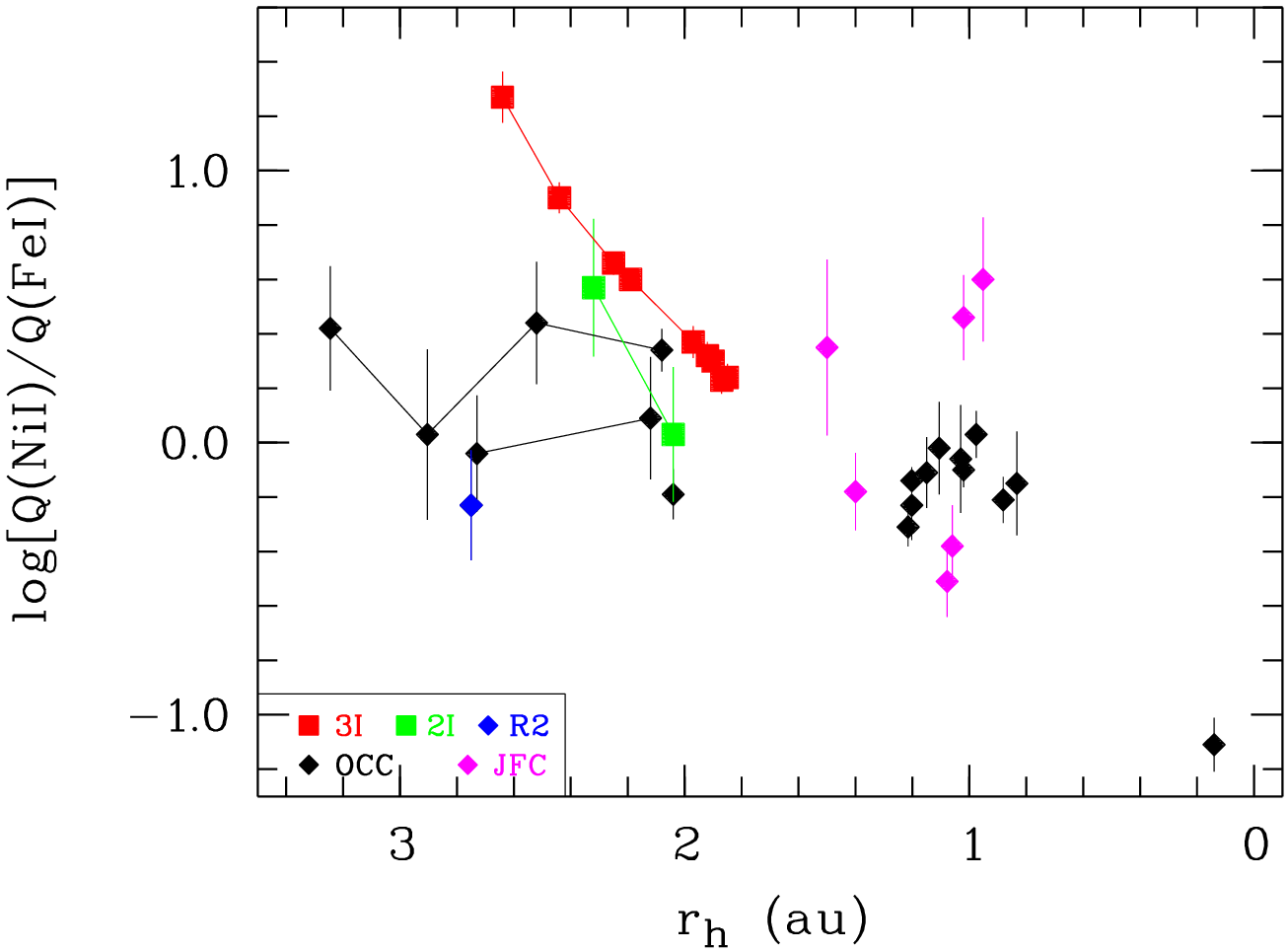}}
\caption{Q(NiI)/Q(FeI) as a function of the heliocentric distance, $r_h$, for interstellar and Solar System comets. The measurements obtained at different heliocentric distances are not averaged and shown individually for comets 3I (9 values, pre-perihelion), 2I (2 values, post-perihelion), C/2001 P1 (4 values, pre-perihelion), and C/2017 K2 (2 values, pre-perihelion).  The comet with the smallest Q(NiI)/Q(FeI) ratio is C/1965 S1(Ikeya-Seki) observed at 0.14 au from the Sun. The solar Ni/Fe abundance ratio is equal, in logarithm,  to $-1.25 \pm 0.04$ \citep{Asplund2009}.}
\label{fig:c05}
\end{figure}

\begin{figure}[]
\centering
\resizebox{0.95\hsize}{!}{\includegraphics*{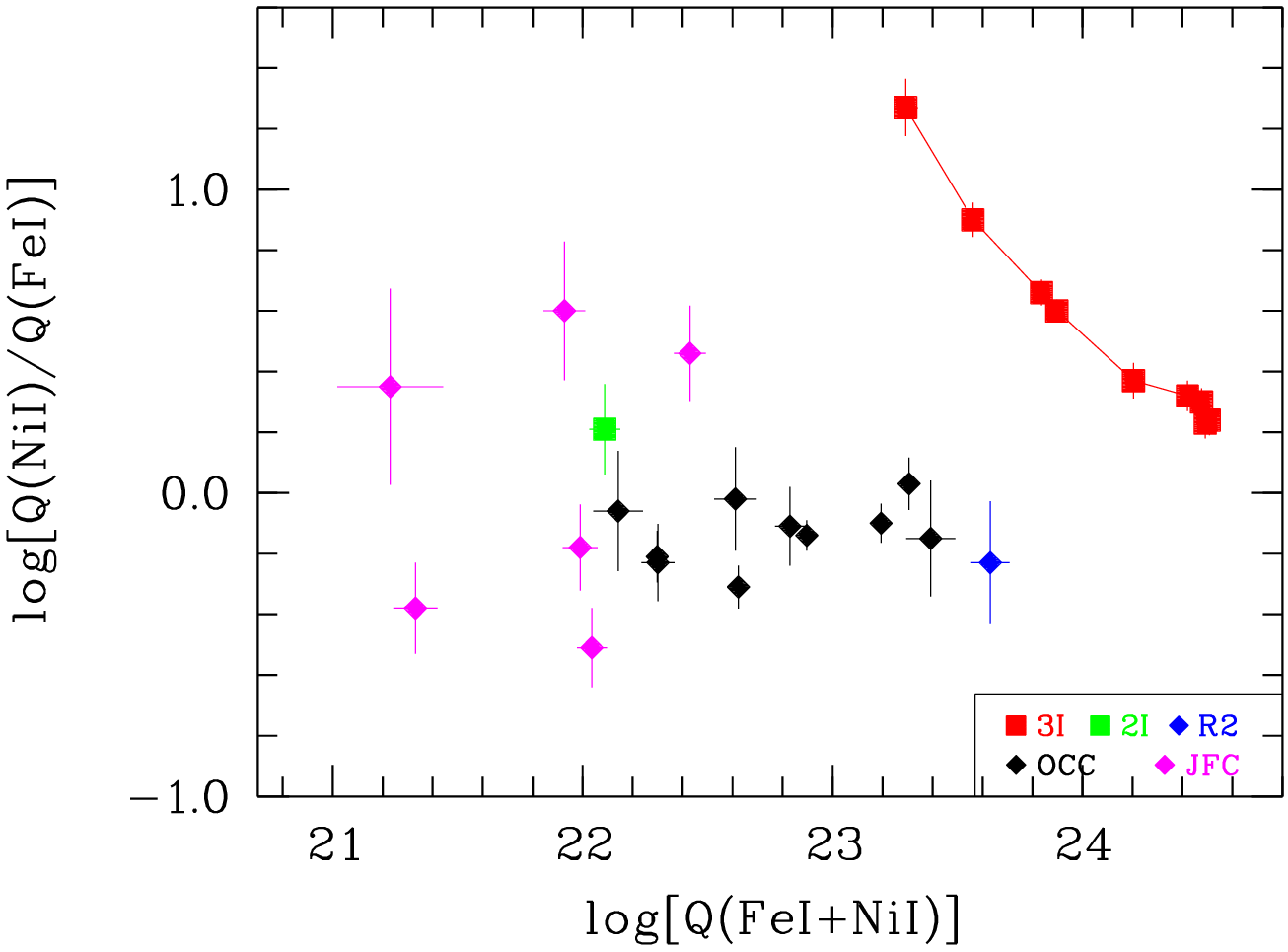}}
\caption{Q(NiI)/Q(FeI) as a function of the total production rate Q(FeI+NiI). Comet 3I was observed in the range 2.64-1.85~au. Solar System comets with $r_h > $ 1.85~au are not shown, except comet C/2016 R2, which was observed at 2.75~au. Comet 2I was observed at 2.18$\pm$0.14~au.}
\label{fig:c25}
\end{figure}

\begin{figure}[]
\centering
\resizebox{0.95\hsize}{!}{\includegraphics*{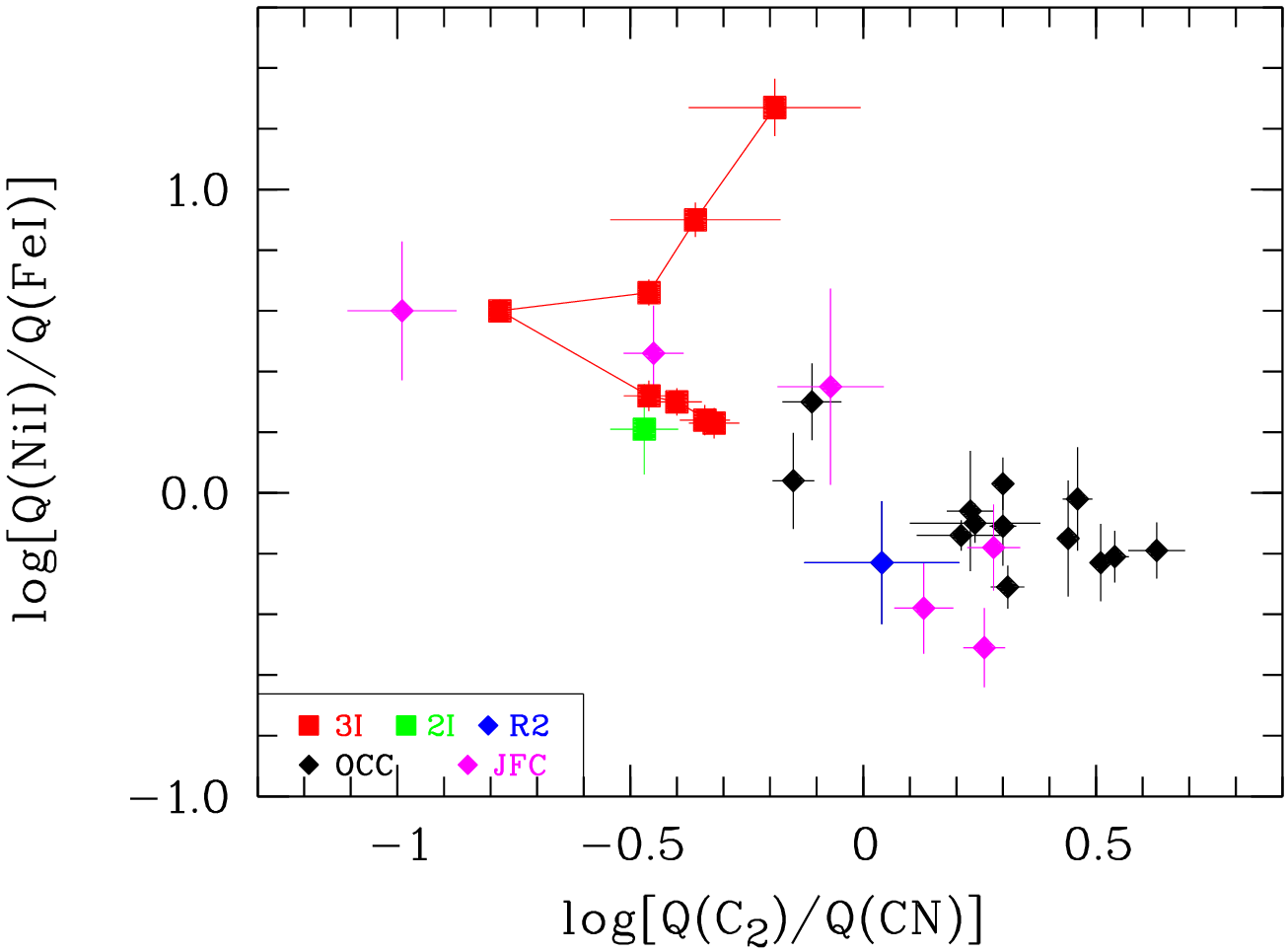}}
\caption{Q(NiI)/Q(FeI) as a function of the production rate ratio Q(C$_2$)/Q(CN).}
\label{fig:c33}
\end{figure}

We compared the NiI and FeI production rates and the  NiI/FeI abundance ratio measured in comet 3I to the same quantities evaluated in other comets (Figs.~\ref{fig:c05} to \ref{fig:c33}). The NiI and FeI production rates were measured in approximately 20 comets based on UVES observations (except comet C/1996 B2 Hyakutake). The data are taken from \citet{Manfroid2021} and \citet{Hutsemekers2021}, as are the production rates of C$_2$ and CN.  These measurements include those obtained for the interstellar comet 2I \citep{Opitom2021}. We also added the data of comet C/2017~K2, which was recently investigated by \citet{Hmiddouch2025}. Unless otherwise stated, when more than one measurement is available for a given comet, we used the simple average, with errors computed from the individual measurements. For comet 3I, however, we show the individual measurements separately. Solar System comets were classified into two broad categories: Jupiter-family comets (JFCs) and Oort-cloud comets (OCCs). For subgroup definitions and the identification of the individual comets in the plots, we refer to \citet{Manfroid2021} and \citet{Hutsemekers2021}. In the plots, we emphasize the measurements of the interstellar comet 2I and of the unusual N$_2$-rich Solar System comet C/2016 R2 \citep{Opitom2019}.

Figure~\ref{fig:c05} shows Q(NiI)/Q(FeI) as a function of the comet heliocentric distance. No correlation was found between Q(NiI)/Q(FeI) and $r_h$ for the Solar System comets \citep{Manfroid2021}. For some comets, Q(NiI)/Q(FeI) was measured at different heliocentric distances, higher than 3~au for C/2000 P1, and it was found to be constant within the error bars.  However, comet 3I breaks the rule by showing an extremely high Q(NiI)/Q(FeI) ratio at $r_h \simeq$ 2.64~au and a rapid  decrease with decreasing $r_h$. After one month, at a heliocentric distance of around 2~au, its value became comparable to that of the other Solar System comets. The unexpected behavior of Q(NiI)/Q(FeI) in comet 3I can be clearly seen thanks to the small errors of the measurements. Therefore, we cannot exclude the possibility that other comets exhibited the same behavior, but were simply not regularly monitored with a signal-to-noise ratio good enough to detect it. The two Q(NiI)/Q(FeI) measurements in the interstellar comet 2I seem to show a trend with the heliocentric distance. However, the ratios are equal within the error bars, so the inferred trend is not statistically significant. Therefore, we only considered the average value of Q(NiI)/Q(FeI) for comet~2I in the following.

Figure~\ref{fig:c25} shows the Q(NiI)/Q(FeI) ratio as a function of the total production rate of the NiI and FeI atoms. On average, OCCs have higher Q(FeI+NiI) rates than JFCs, while the latter show a larger dispersion of the Q(NiI)/Q(FeI) ratio \citep{Hutsemekers2021}. In this plot, comet 3I is clearly exceptional. When the Q(NiI)/Q(FeI) ratio becomes similar to that observed in the other comets, the total Q(FeI+NiI) production rate in comet 3I is one order of magnitude larger than the highest value measured in Solar System comets at heliocentric distance smaller than 1.85~au. Comet 2I, on the other hand, is just ordinary, with a total Q(FeI+NiI) production rate two orders of magnitude smaller than that of comet 3I, at a comparable heliocentric distance. It is also interesting to note that, at $r_h \simeq$ 2.75~au, Q(FeI+NiI) in comet C/2016 R2 was a factor of about 2.5 times larger than in comet 3I. However, unlike comet 3I, the metal production in comet C/2016 R2 was dominated by FeI and not NiI. Unfortunately, comet C/2016 R2 was not observed at other heliocentric distances.

Figure~\ref{fig:c33} shows the correlation between Q(NiI)/Q(FeI) and Q(C$_2$)/Q(CN) previously established for the Solar System comets \citep{Hutsemekers2021}. Given the variation of Q(NiI)/Q(FeI) in comet 3I, one might thus wonder how it fits the relationship. Ignoring the leftmost data point, where Q(CN) temporarily increases faster than Q(C2), $\log$[Q(C$_2$)/Q(CN)] for comet 3I consistently lies between $-$0.5 and $-$0.3 at all heliocentric distances (epochs). With $\log$[Q(C$_2$)/Q(CN)] $\lesssim  -0.13$ \citep{Bair2025}, comet 3I can thus be confidently classified as a C$_2$-depleted comet, though not as strongly depleted as is reported by \citet{Salazar2025}. Comet 3I perfectly fits the correlation between Q(NiI)/Q(FeI) and Q(C$_2$)/Q(CN) once $\log$[Q(NiI)/Q(FeI)] stabilizes around 0.3 when the comet was at $r_h \simeq$ 2~au. While within this correlation, the properties of comet 3I appear very similar to those of comet 2I. 

It is also interesting to note that, between $r_h$ = 2.64 and 2.19 au, $\log$[Q(CN)/Q(OH)] in comet 3I is in the range of $-1.8$ to $-2.2$. These values are consistent with those measured for Solar System comets \citep{AHearn1995}, though they are at the higher end. However, our measurements are significantly higher than the ratio of $\log$[Q(CN)/Q(OH)] $\simeq -3.1$ reported by \citet{Xing2025} when comet 3I was at 2.9~au. 

In summary, comet 3I exhibits an extremely high Q(NiI)/Q(FeI) ratio at the onset of FeI. Later, this ratio decreases regularly and, at $r_h \simeq$ 2~au, becomes indistinguishable from the ratio measured in other Solar System comets and the interstellar comet 2I. On the other hand, the total production rate of NiI and FeI atoms, which was not exceptional at $r_h \simeq $ 3~au,  strongly increases and, at $r_h \simeq$ 2~au, becomes more than one order of magnitude larger than that of other comets observed at comparable or smaller heliocentric distances.

\section{Discussion: The origin of NiI and FeI atoms}
\label{sec:discu}

The expected blackbody equilibrium temperature, $T$, (in Kelvin) of the cometary surface is approximately given by
\begin{equation}
\label{eq:t}
T \simeq 280 \; r_{h}^{-1/2} \; ,
\end{equation}
where $r_{h}$ is in astronomical units \citep{Manfroid2021,Guzik2021,Puzia2025}. At the distances at which the comets were observed, this temperature is far too low to vaporize silicate, sulfide, and metallic grains that contain NiI and FeI atoms. Therefore, the presence of NiI and FeI atoms in cometary comae is extremely puzzling. Several hypotheses have been proposed to explain the presence of the NiI and FeI atoms at low surface temperatures as well as the high NiI/FeI abundance ratio, much higher that the Solar System value. These scenarios were first presented by \citet{Manfroid2021}, and discussed further by \citet{Bromley2021} and \citet{Rahatgaonkar2025}. In particular, \citet{Manfroid2021} proposed that the NiI and FeI atoms could be released from  Ni(CO)$_4$ and Fe(CO)$_5$ carbonyls, which are characterized by very low sublimation temperatures. The smaller sublimation temperature of Ni(CO)$_4$ with respect to Fe(CO)$_5$ could also explain the high NiI/FeI ratios observed at low surface temperatures. The carbonyl scenario, although appealing, predicts a NiI/FeI ratio that strongly depends on the temperature, and thus on the heliocentric distance. Unfortunately, such a dependence was not observed in the sample of Solar System comets studied by \citet{Manfroid2021}.

On the contrary, comet 3I exhibits significant variation in its NiI/FeI abundance ratio as a function of the heliocentric distance, enabling us to further explore this scenario. \citet{Rahatgaonkar2025} already noticed that the production rate of NiI changes much faster with the heliocentric distance than can be explained by purely radiative processes.

\citet{Bromley2021} showed that the photodissociation of Ni(CO)$_4$ and Fe(CO)$_5$ into atomic NiI and FeI is very fast, with similar rates for both carbonyls, and occurs at nucleocentric distances consistent with the observed spatial distributions of the atoms. Therefore, the NiI and FeI production rates can be directly compared to the carbonyl sublimation rates. The sublimation rates of  Ni(CO)$_4$ and Fe(CO)$_5$ were computed as a function of the temperature as in \citet{Manfroid2021}. Then, they were expressed as a function of $r_{h}$ using Eq.~\ref{eq:t}.  The ratio of the sublimation rates was finally corrected by assuming an intrinsic cometary Ni/Fe abundance ratio equal to the Solar System Ni/Fe abundance ratio\footnote{It should be emphasized that the Ni/Fe abundance ratio in the cometary material (ices and dust) differs from the abundance ratio of the NiI and FeI atoms in the coma, which depends on the sublimation~rates.}, $\log_{10}$ (Ni/Fe)  = $-1.25$ \citep{Asplund2009}. 

\begin{figure}[]
\centering
\resizebox{0.95\hsize}{!}{\includegraphics*{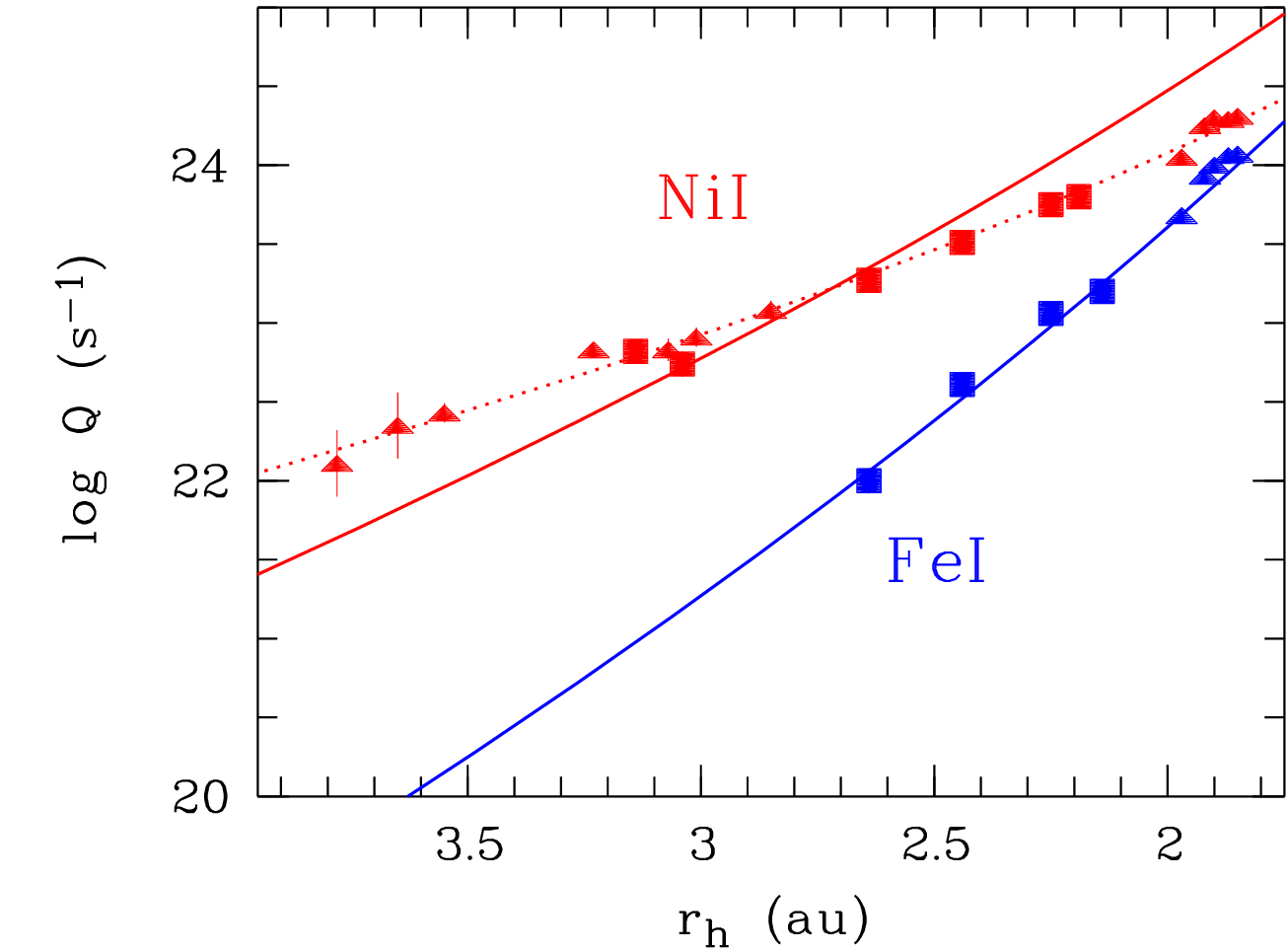}}
\caption{Q(NiI) and Q(FeI) production rates in comet 3I as a function of the heliocentric distance. Squares represent UVES measurements and triangles X-shooter measurements. The curves show the production rates of NiI (in red) and FeI (in blue), which were obtained by multiplying the sublimation rates of the Ni(CO)$_4$ and  Fe(CO)$_5$ carbonyls by the effective emitting area of each species (see text). For the solid curves, the sublimation rates were computed assuming $T = 280 \; r_{h}^{-1/2}$. For the dotted red curve, the sublimation rate was computed assuming $T = 237 \; r_{h}^{-1/3}$ for Ni(CO)$_4$. } 
\label{fig:zqni}
\end{figure}

Figure~\ref{fig:zqni} shows the variation in Q(NiI) between 3.78 and 1.85~au, and the variation in Q(FeI) between 2.64~au and 1.85~au. The solid curves represent the production rates of NiI and FeI computed from the sublimation rates of Ni(CO)$_4$ and Fe(CO)$_5$, which were multiplied by the effective emitting area of each species. The sublimation rates were computed assuming that the comet was at the equilibrium temperature at each heliocentric distance (Eq.~\ref{eq:t}). The effective emitting area was estimated empirically  by shifting vertically (in logarithm) the sublimation rate curves (in molecules m$^{-2}$ s$^{-1}$) to match the observed production rates (in atoms s$^{-1}$). The effective emitting areas were found to be equal to 1.6 m$^2$ for both Ni(CO)$_4$ and Fe(CO)$_5$.  This area is small, as is expected for minor species. However, its value should be interpreted with caution given the basic model. The Fe(CO)$_5$ sublimation model nicely reproduces the variation in Q(FeI) with the heliocentric distance. For NiI, the global trend is reproduced, indicating that carbonyl sublimation begins very early. However, the increase in Q(NiI) with the heliocentric distance is less steep than the simple sublimation model predicts. This suggests that the temperature of the NiI-emitting regions may differ from the blackbody equilibrium temperature. Indeed, a reasonable fit can be obtained using the ad hoc relation $T = 237 \; r_{h}^{-1/3}$. As the comet approaches the Sun, more material is ejected, fractures form, and the temperature can deviate significantly from the blackbody equilibrium temperature, at least in localized regions \citep[e.g.,][]{Prialnik2004,Groussin2013,Hofner2017}. The different behavior of NiI and FeI with respect to the simple sublimation model could also indicate that these atoms are partially released from different subregions.

\begin{figure}[]
\centering
\resizebox{0.95\hsize}{!}{\includegraphics*{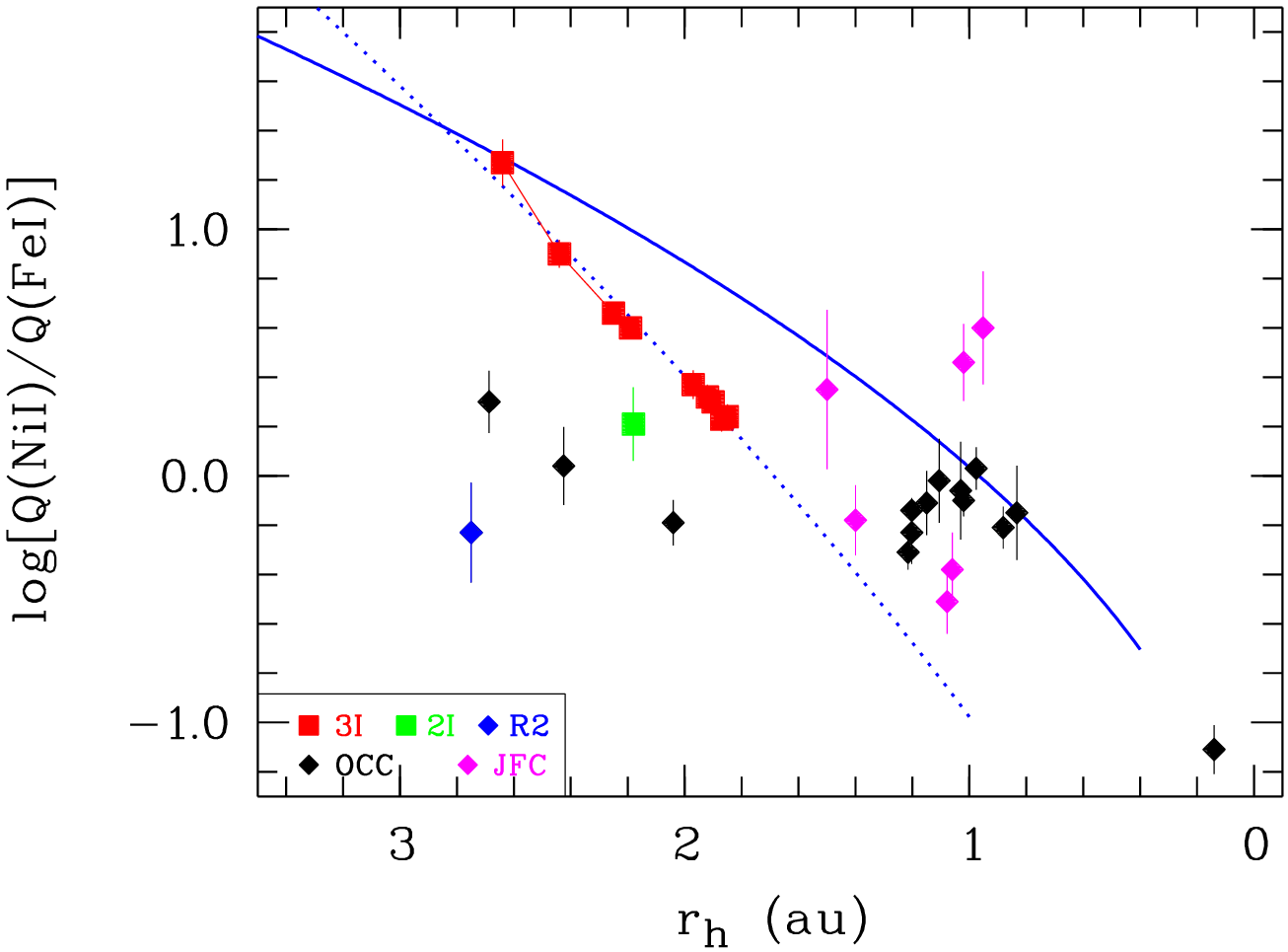}}
\caption{Q(NiI)/Q(FeI) as a function of the heliocentric distance for interstellar and Solar System comets. The solid curve represents the ratio of the sublimation rates of the Ni(CO)$_4$ and Fe(CO)$_5$ carbonyls computed assuming $T = 280 \; r_{h}^{-1/2} $ for both Ni(CO)$_4$ and Fe(CO)$_5$. The dotted line represents the sublimation rate ratio assuming $T = 237 \; r_{h}^{-1/3}$ for the Ni(CO)$_4$ emitting region, as in Fig.~\ref{fig:zqni}.} 
\label{fig:c06}
\end{figure}

Figure~\ref{fig:c06} shows the ratio of the sublimation rates of Ni(CO)$_4$ and Fe(CO)$_5$ computed as a function of $r_{h}$ according to Eq.~\ref{eq:t}, together with the NiI/FeI abundance ratios measured in interstellar and Solar System comets. As before, the ratio of the sublimation rates is adjusted to account for the cometary intrinsic Ni/Fe abundance ratio, which is assumed to be equal to the Solar System value. Although the model is extremely basic, the computed ratio of the sublimation rates of Ni(CO)$_4$ to Fe(CO)$_5$ (solid curve) reproduces the NiI/FeI ratio observed in several comets within a factor 2-3, though not in all of them. In particular, it accurately reproduces the ratio observed in comet 3I at the onset of FeI at $r_h \simeq$ 2.64 au. However, the rapid decrease in the NiI/FeI ratio with $r_h$  does not follow the theoretical curve computed under the hypothesis that the comet remains at the equilibrium temperature. Assuming a different temperature dependence, such as $T = 237 \; r_{h}^{-1/3}$, allows us to reproduce the NiI/FeI variation (dotted curve). Therefore, the ratio of the carbonyl sublimation rates at the equilibrium temperature should be considered as an upper limit of the observed NiI/FeI ratios.  The lack of correlation between the NiI/FeI ratio and $r_{h}$ for Solar System comets may be due to the dispersion of their temperatures and surface properties at a given heliocentric distance.

Although incomplete, this scenario provides a straightforward explanation for the release of NiI and FeI atoms in comets and suggests the presence of carbonyls in the cometary material. However, it is clear that there is more to the story. The NiI/FeI ratio is also connected to the carbon content, as is shown by the relationship between the NiI/FeI and C$_2$/CN ratios (Fig.~\ref{fig:c33}). Since the C$_2$ depletion is primordial rather than evolutionary \citep{Bair2025}, the NiI/FeI abundance ratio could also be primordial in part. Comet 3I exhibits a CO$_2$/H$_2$O abundance ratio of 7.6 at $r_h$ = 3.3~au, which is higher than that of Solar System comets, and a CO/H$_2$O ratio of 1.7, which is consistent with other cometary observations \citep{Cordiner2025}, including with those of comet 2I, which exhibits CO/H$_2$O ratios ranging from 1.3 to 1.6 at 2.0~au \citep{Bodewits2020}. Comet C/2016 R2 exhibits more extreme abundance ratios:  CO$_2$/H$_2$O $\simeq$ 30 and CO/H$_2$O $\simeq$ 300 at $r_h$ = 2.8~au \citep{McKay2019}. The high Q(FeI+NiI) production rates observed in comets 3I and C/2016 R2 at approximately 2-3~au (Sect.~\ref{sec:results} and Fig.~\ref{fig:c25}) could be related to the high abundance of CO$_2$, CO, or both. As is suggested by \citet{Cordiner2025}, the high CO$_2$/H$_2$O measured in comet 3I could be due to a lower temperature of the cometary surface, which would also affect the metal production. On the other hand, when more CO$_2$ and CO molecules are available in the cometary material, more NiI and FeI atoms could be incorporated into carbonyls during the comet formation, so that more NiI and FeI atoms are finally produced when these carbonyls sublimate. A better understanding of carbonyl formation in various chemical environments is needed, in particular to determine whether an initial Ni(CO)$_4$ / Fe(CO)$_5$ asymmetry can already be produced during the comet formation, depending on the carbon oxide content. Interestingly, Q(FeI+NiI) is dominated by NiI in comet 3I (Fig.~\ref{fig:c25}), which has a high CO$_2$/CO ratio, while Q(FeI+NiI) is dominated by FeI in comet C/2016 R2, which, on the contrary, has a high CO/CO$_2$ ratio. Finally, it is worth noting that other mechanisms, such as superheating of nanograins, could also contribute to the release of NiI and FeI atoms from refractory material. However such mechanisms cannot fully explain the observed NiI/FeI ratios \citep[see][]{Manfroid2021}. For Sun-grazing comets, complete vaporization of refractory grains occurs, leading to a solar NiI/FeI abundance ratio.

\section{Conclusions}
\label{sec:conclu}

Since the beginning of its activity, the interstellar comet 3I/ATLAS has been regularly observed with UVES and X-shooter. These observations have provided an unprecedented dataset revealing the evolution of the NiI and FeI emission lines with heliocentric distance. During the initial stages of its activity, comet 3I exhibited extreme and unusual NiI/FeI abundance ratios. However, as its heliocentric distance decreased, the ratio became indistinguishable from those observed in Solar System comets and in the interstellar comet 2I/Borisov. Comet 3I was found to be C$_2$-depleted, with a NiI/FeI abundance ratio finally consistent with other C$_2$-depleted comets. Nevertheless, comet 3I remains exceptional due to its high total production rate of NiI and FeI, which is at least one order of magnitude larger than that of other comets. This enhanced metal production is possibly related to other chemical anomalies, such as the high CO$_2$/H$_2$O  abundance ratio.  

The NiI and FeI production rates, as well as their variations with heliocentric distance, are interpreted within the context of the sublimation of Ni(CO)$_4$ and Fe(CO)$_5$ carbonyls. This scenario provides a straightforward explanation for the asymmetric release of NiI and FeI atoms in the cometary coma and how it depends on heliocentric distance. Post-perihelion observations of NiI and FeI would be helpful to confirm the findings of this paper and clarify the trend with the heliocentric distance, in particular to discriminate between genuine variation with the heliocentric distance and effects due to processing of the outer layers of the nucleus.

\begin{acknowledgements}
We thank the referee for comments that helped significantly improve the manuscript. DH and EJ are Research Directors at the F.R.S-FNRS. JM is honorary Research Director at the F.R.S-FNRS. THP, RR, JPC, BL gratefully acknowledge support from the National Agency for Research and Development (ANID) under the following grants: CATA-Basal FB210003 and Beca de Doctorado Nacional (R.R. and J.P.C.). R.C.D. acknowledges support from grant \#361233 awarded by the Research Council of Finland to M. Granvik.
\end{acknowledgements}

\bibliographystyle{aa}
\bibliography{references}

\onecolumn

\begin{appendix}
\section{Tables}
\label{sec:appendixA}

\begin{table*}[h]
\caption{Observing circumstances: UVES.}
\label{tab:obs1}
\centering
\begin{tabular}{lccccccc}
\hline\hline
 Date       & $r_h$ &  $\dot{r_h}$ & $\Delta$ & $\dot{\Delta}$ & Settings & $w_B / w_R$ & $h_B / h_R$ \\
 yyyy-mm-dd & au    &  km s$^{-1}$ & au        & km s$^{-1}$    &             & \arcsec & \arcsec \\
\hline 
2025-08-12  & 3.14 & -55.3  &  2.69  & -15.4 &  346+580   &  1.8/0.6 &  9.5/11.5   \\
2025-08-15  & 3.04 & -54.9  &  2.66  & -13.3 &  390+580   &  0.6/0.6 &  7.5/11.5   \\
2025-08-28  & 2.64 & -52.9  &  2.59  &  -6.6 &  348+580   &  1.8/0.6 &  9.5/11.5   \\
2025-08-28  & 2.64 & -52.9  &  2.59  &  -6.6 &  437+860   &  1.8/0.6 &  9.5/11.0   \\
2025-09-03  & 2.46 & -51.6  &  2.57  &  -4.8 &  348+580   &  1.8/1.2 &  9.5/11.5   \\
2025-09-03  & 2.46 & -51.6  &  2.57  &  -4.8 &  437+860   &  1.8/1.2 &  9.5/11.0   \\
2025-09-04  & 2.43 & -51.3  &  2.57  &  -4.6 &  348+580   &  1.8/1.2 &  9.5/11.5   \\
2025-09-04  & 2.43 & -51.3  &  2.57  &  -4.6 &  437+860   &  1.8/1.2 &  9.5/11.0   \\
2025-09-10  & 2.25 & -49.5  &  2.55  &  -3.7 &  348+580   &  1.8/1.2 &  9.5/11.5   \\
2025-09-11  & 2.22 & -49.2  &  2.55  &  -3.7 &  437+860   &  1.8/1.2 &  9.5/11.0   \\
2025-09-12  & 2.19 & -48.8  &  2.55  &  -3.6 &  348+580   &  1.8/1.2 &  9.5/11.5   \\
2025-09-14  & 2.14 & -48.1  &  2.54  &  -3.6 &  437+860   &  1.8/1.2 &  9.5/11.0   \\
\hline
\end{tabular}
\end{table*}

\begin{table*}[h]
\caption{Observing circumstances: X-shooter.}
\label{tab:obs2}
\centering
\begin{tabular}{lccccc}
\hline\hline
 Date       & $r_h$ &  $\dot{r_h}$ & $\Delta$ & $\dot{\Delta}$ & $w_B / h_B$ \\
 yyyy-mm-dd & au    &  km s$^{-1}$ & au        & km s$^{-1}$   & \arcsec \\ 
\hline 
2025-07-23  & 3.78 & -56.8  &  2.96  & -32.5  &  1.6 /11.0 \\
2025-07-27  & 3.65 & -56.6  &  2.89  & -28.7  &  1.6 /11.0\\
2025-07-30  & 3.55 & -56.4  &  2.84  & -26.0  &  1.6 /11.0 \\
2025-08-09  & 3.23 & -55.6  &  2.72  & -17.6  &  1.6 /11.0\\
2025-08-14  & 3.07 & -55.0  &  2.67  & -14.0  &  1.6 /11.0\\
2025-08-16  & 3.01 & -54.8  &  2.65  & -12.7  &  1.6 /11.0 \\
2025-08-21  & 2.85 & -54.1  &  2.62  &  -9.7  &  1.6 /11.0\\
2025-09-20  & 1.97 & -45.4  &  2.53  &  -3.9  &  1.6 /11.0\\
2025-09-22  & 1.92 & -44.3  &  2.52  &  -4.2  &  1.6 /11.0\\
2025-09-23  & 1.90 & -43.7  &  2.52  &  -4.4  &  1.6 /11.0\\
2025-09-24  & 1.87 & -43.1  &  2.52  &  -4.5  &  1.6 /11.0\\
2025-09-25  & 1.85 & -42.5  &  2.52  &  -4.7  &  1.6 /11.0\\
\hline
\end{tabular}
\tablefoot{In Tables \ref{tab:obs1} and \ref{tab:obs2}, $r_h$ and $\Delta$ are the heliocentric and geocentric distances of the comet. $\dot{r_h}$ and $\dot{\Delta}$ are the corresponding velocities. $w_B$, $w_R$, $h_B$, and $h_R$  refer to the blue/red slit width and height, respectively. }
\end{table*}

\begin{table*}[h]
\caption{FeI and NiI production rates, and their ratio.}
\label{tab:data1}
\centering
\begin{tabular}{lccccccccccc}
\hline\hline
 Date & $r_h$ & n$_{\rm lines}$  & log$_{10}$ Q(FeI) & log$_{10}$ Q(NiI) & log$_{10}$ [Q(NiI)/Q(FeI)] & Instrument \\
 yyyy-mm-dd & au     &  FeI / NiI  &   s$^{-1}$     &   s$^{-1}$ &        &     \\
\hline 
2025-07-23  & 3.78 &    - /  2   &   -                & 22.11 $\pm$ 0.21  & -                & X-shooter   \\
2025-07-27  & 3.65 &    - /  2   &   -                & 22.35 $\pm$ 0.21  & -                & X-shooter   \\
2025-07-30  & 3.55 &    - /  4   &   -                & 22.43 $\pm$ 0.06  & -                & X-shooter   \\
2025-08-09  & 3.23 &    - /  3   &   -                & 22.83 $\pm$ 0.02  & -                & X-shooter   \\
2025-08-12  & 3.14 &    - / 20   & $<$ 21.83          & 22.82 $\pm$ 0.06  &   $>$ 0.99       & UVES \\
2025-08-14  & 3.07 &    - / 11   &   -                & 22.83 $\pm$ 0.07  & -                & X-shooter   \\
2025-08-15  & 3.04 &    - / 14   & $<$ 22.14          & 22.74 $\pm$ 0.08  &   $>$ 0.60       & UVES \\
2025-08-16  & 3.01 &    - / 12   &   -                & 22.91 $\pm$ 0.06  & -                & X-shooter   \\
2025-08-21  & 2.85 &    - / 14   &   -                & 23.08 $\pm$ 0.06  & -                & X-shooter   \\
2025-08-28  & 2.64 &    5 / 28   &  22.00 $\pm$ 0.08  & 23.27 $\pm$ 0.05  & 1.27 $\pm$ 0.10  & UVES \\
2025-09-3/4 & 2.44 &   17 / 36   &  22.61 $\pm$ 0.04  & 23.51 $\pm$ 0.04  & 0.90 $\pm$ 0.06  & UVES \\
2025-09-10  & 2.25 &   42 / 36   &  23.06 $\pm$ 0.03  & 23.75 $\pm$ 0.03  & 0.66 $\pm$ 0.04  & UVES \\
2025-09-12  & 2.19 &   46 / 37   &  23.20 $\pm$ 0.02  & 23.80 $\pm$ 0.03  & 0.60 $\pm$ 0.04  & UVES \\
2025-09-20  & 1.97 &   11 / 18   &  23.68 $\pm$ 0.03  & 24.05 $\pm$ 0.05  & 0.37 $\pm$ 0.06  & X-shooter   \\
2025-09-22  & 1.92 &   16 / 18   &  23.93 $\pm$ 0.03  & 24.25 $\pm$ 0.04  & 0.32 $\pm$ 0.05  & X-shooter   \\
2025-09-23  & 1.90 &   21 / 18   &  24.00 $\pm$ 0.02  & 24.30 $\pm$ 0.04  & 0.30 $\pm$ 0.05  & X-shooter   \\
2025-09-24  & 1.87 &   18 / 20   &  24.06 $\pm$ 0.03  & 24.29 $\pm$ 0.04  & 0.23 $\pm$ 0.05  & X-shooter   \\
2025-09-25  & 1.85 &   16 / 20   &  24.07 $\pm$ 0.03  & 24.31 $\pm$ 0.04  & 0.24 $\pm$ 0.05  & X-shooter   \\
\hline
\end{tabular}
\end{table*}

\begin{table*}[h]
\caption{OH, CN, and C$_2$ production rates.}
\label{tab:data2}
\centering
\begin{tabular}{lccccc}
\hline\hline
 Date            & $r_h$  & log$_{10}$ Q(OH) & log$_{10}$ Q(CN) & log$_{10}$ Q(C$_2$) & Instrument \\
 yyyy-mm-dd      & au     &   s$^{-1}$      &   s$^{-1}$       &   s$^{-1}$          &           \\
 \hline
2025-08-12  & 3.14 &  -                & 24.2  $\pm$ 0.3  &    -                & UVES \\
2025-08-15  & 3.04 &  -                & 24.59 $\pm$ 0.05 &    -                & UVES  \\
2025-08-28  & 2.64 &  26.58 $\pm$ 0.04 & 24.79 $\pm$ 0.04 &  24.6 $\pm$ 0.2     & UVES \\
2025-09-3/4 & 2.44 &  26.79 $\pm$ 0.02 & 24.96 $\pm$ 0.03 &  24.6 $\pm$ 0.2     & UVES \\
2025-09-10  & 2.25 &  27.24 $\pm$ 0.13 & 25.03 $\pm$ 0.01 &  24.57 $\pm$ 0.02   & UVES \\
2025-09-11  & 2.22 &  -                & 25.34 $\pm$ 0.01 & -                   & UVES \\
2025-09-12  & 2.19 &  27.33 $\pm$ 0.10 & 25.46 $\pm$ 0.01 &  24.68 $\pm$ 0.01   & UVES \\
2025-09-14  & 2.14 &  -                & 25.43 $\pm$ 0.01 & -                   & UVES \\
2025-09-20  & 1.97 &  -                & 25.61 $\pm$ 0.02 & -                & X-shooter  \\
2025-09-22  & 1.92 &  -                & 25.70 $\pm$ 0.02 & 25.24 $\pm$ 0.05 & X-shooter  \\
2025-09-23  & 1.90 &  -                & 25.72 $\pm$ 0.02 & 25.32 $\pm$ 0.05 & X-shooter  \\
2025-09-24  & 1.87 &  -                & 25.73 $\pm$ 0.02 & 25.41 $\pm$ 0.05 & X-shooter  \\
2025-09-25  & 1.85 &  -                & 25.73 $\pm$ 0.02 & 25.39 $\pm$ 0.05 & X-shooter  \\
\hline
\end{tabular}
\tablefoot{In Tables \ref{tab:data1} and \ref{tab:data2}, the spectra obtained on September~3 and~4 in very similar circumstances were averaged to increase the signal-to-noise ratio. OH production rates were only measured in the UVES spectra. CN production rates before August 21, 2025 can be found in \citet{Rahatgaonkar2025}. }
\end{table*}

\clearpage

\section{Evolution of NiI and FeI emission lines}
\label{sec:appendixB}

\begin{figure*}[h]
\centering
\resizebox{0.75\hsize}{!}{\includegraphics*{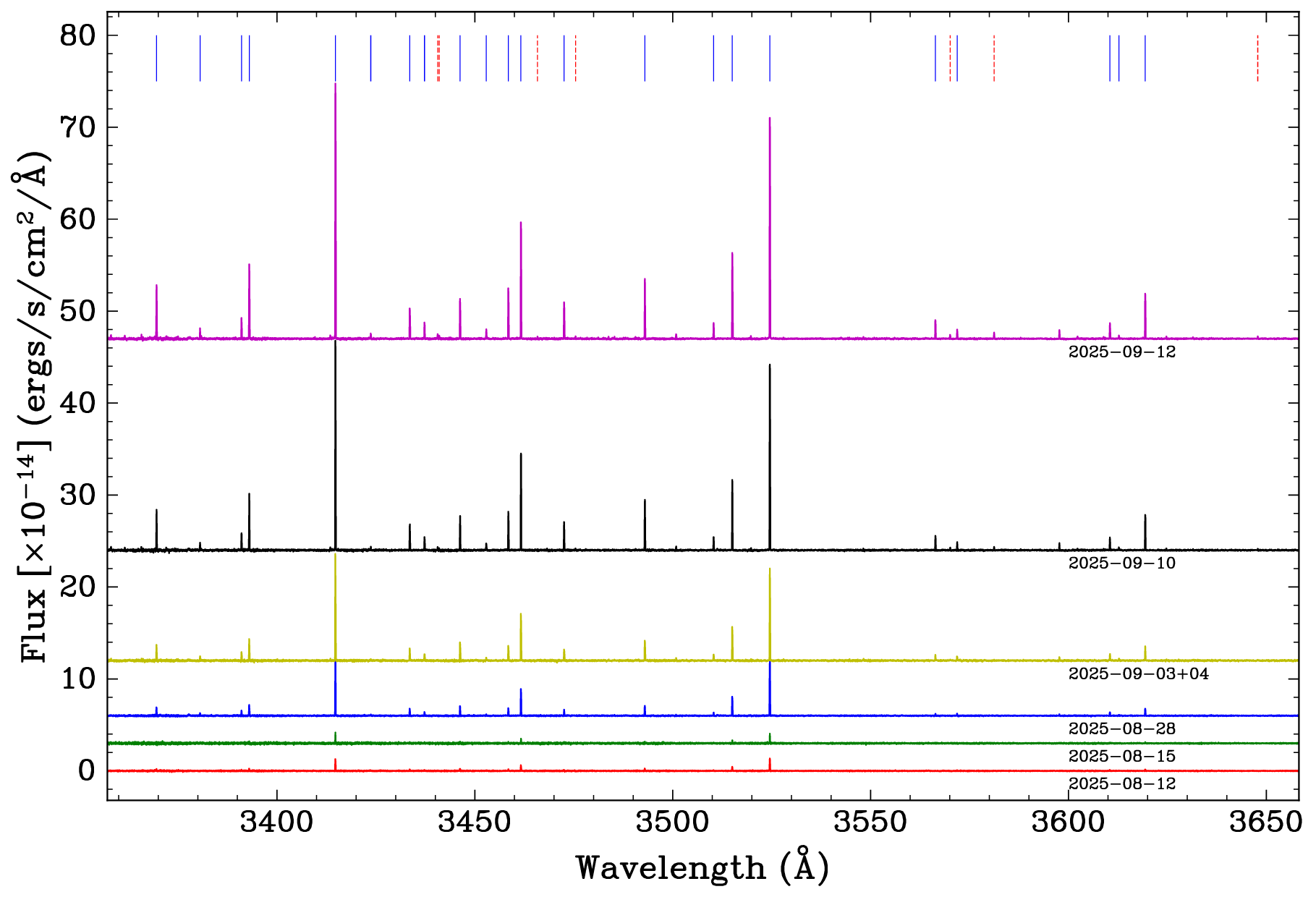}}
\caption{UVES blue, continuum-subtracted, spectra of comet 3I from August 12 to September 12. They show the evolution of the NiI (solid blue tickmarks) and FeI lines (dashed red tickmarks) in the spectral range 3360-3650 \AA. The flux scale is identical for all spectra, which are shifted vertically for clarity.}  
\label{fig:spec_ni}
\end{figure*}
\begin{figure*}[]
\centering
\resizebox{0.75\hsize}{!}{\includegraphics*{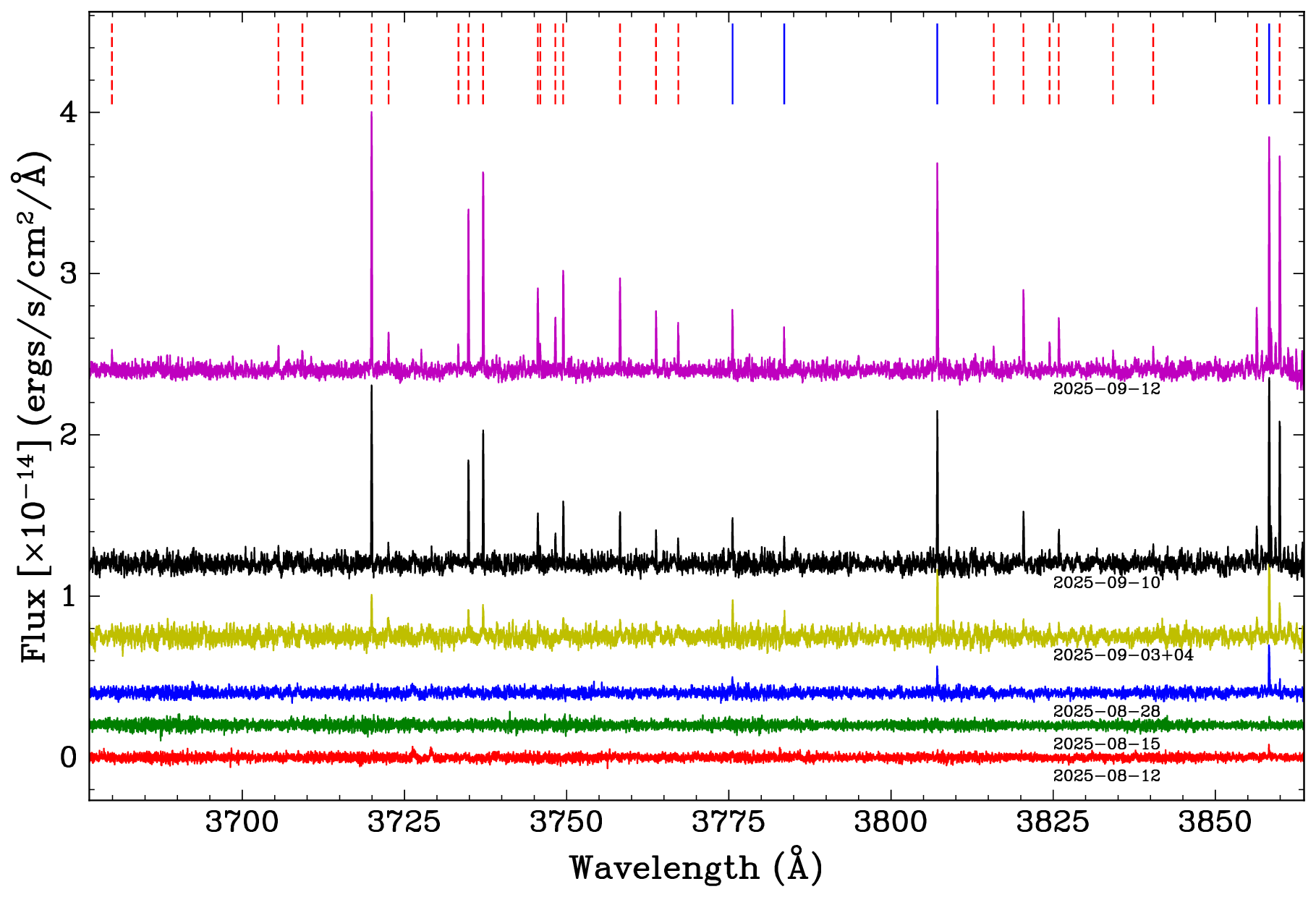}}
\caption{Same as Fig.~\ref{fig:spec_ni} for the 3680 - 3860 \AA\ spectral range.}    
\label{fig:spec_fe}
\end{figure*}

\end{appendix}

\end{document}